\pdfoutput=1
\documentclass[12pt,a4paper]{article}

\usepackage{ifthen} % for conditional statements
\newboolean{pdflatex}
\setboolean{pdflatex}{true} % False for eps figures 

\newboolean{articletitles}
\setboolean{articletitles}{true} % False removes titles in references

\newboolean{uprightparticles}
\setboolean{uprightparticles}{false} %True for upright particle symbols

\newboolean{inbibliography}
\setboolean{inbibliography}{false} %True once you enter the bibliography

\usepackage{microtype}
\usepackage{lineno}  % for line numbering during review
\usepackage{xspace} % To avoid problems with missing or double spaces after
\usepackage{caption} %these three command get the figure and table captions automatically small

\usepackage{graphicx}  % to include figures (can also use other packages)
\usepackage{color}
\usepackage{colortbl}
\graphicspath{{./figs/}} % Make Latex search fig subdir for figures
\usepackage{booktabs}

\usepackage{amsmath} % Adds a large collection of math symbols
\usepackage{amssymb}
\usepackage{amsfonts}
\usepackage{upgreek} % Adds in support for greek letters in roman typeset

\usepackage{hyperref}    % Hyperlinks in references
\usepackage[all]{hypcap} % Internal hyperlinks to floats.

\def\lhcb {\mbox{LHCb}\xspace}

\def\MagUp {\mbox{\em Mag\kern -0.05em Up}\xspace}

\ifthenelse{\boolean{uprightparticles}}%
{

 \def\Pmu         {\ensuremath{\upmu}\xspace}

 \def\Ppi         {\ensuremath{\uppi}\xspace}

 \def\Ppsi        {\ensuremath{\uppsi}\xspace}

 \def\PDelta      {\ensuremath{\Delta}\xspace}                 
 \def\PXi      {\ensuremath{\Xi}\xspace}                 
 \def\PLambda      {\ensuremath{\Lambda}\xspace}                 
 \def\PSigma      {\ensuremath{\Sigma}\xspace}                 
 \def\POmega      {\ensuremath{\Omega}\xspace}                 
 \def\PUpsilon      {\ensuremath{\Upsilon}\xspace}                 
 
 \def\PB      {\ensuremath{\mathrm{B}}\xspace}                 
                  
 \def\PD      {\ensuremath{\mathrm{D}}\xspace}

 \def\PJ      {\ensuremath{\mathrm{J}}\xspace}                 
 \def\PK      {\ensuremath{\mathrm{K}}\xspace}

 \def\Pb      {\ensuremath{\mathrm{b}}\xspace}                 
 \def\Pc      {\ensuremath{\mathrm{c}}\xspace}                 
 \def\Pd      {\ensuremath{\mathrm{d}}\xspace}

 \def\Pi      {\ensuremath{\mathrm{i}}\xspace}

 \def\Pp      {\ensuremath{\mathrm{p}}\xspace}

 \def\Ps      {\ensuremath{\mathrm{s}}\xspace}

}
{

 \def\Pmu         {\ensuremath{\mu}\xspace}

 \def\Ppi         {\ensuremath{\pi}\xspace}

 \def\Ppsi        {\ensuremath{\psi}\xspace}                 
                  
 \mathchardef\PDelta="7101
 \mathchardef\PXi="7104
 \mathchardef\PLambda="7103
 \mathchardef\PSigma="7106
 \mathchardef\POmega="710A
 \mathchardef\PUpsilon="7107
                  
 \def\PB      {\ensuremath{B}\xspace}                 
                  
 \def\PD      {\ensuremath{D}\xspace}

 \def\PJ      {\ensuremath{J}\xspace}                 
 \def\PK      {\ensuremath{K}\xspace}

 \def\Pb      {\ensuremath{b}\xspace}                 
 \def\Pc      {\ensuremath{c}\xspace}                 
 \def\Pd      {\ensuremath{d}\xspace}

 \def\Pi      {\ensuremath{i}\xspace}

 \def\Pp      {\ensuremath{p}\xspace}

 \def\Ps      {\ensuremath{s}\xspace}

}

\makeatletter
\ifcase \@ptsize \relax% 10pt
  \newcommand{\miniscule}{\@setfontsize\miniscule{4}{5}}% \tiny: 5/6
\or% 11pt
  \newcommand{\miniscule}{\@setfontsize\miniscule{5}{6}}% \tiny: 6/7
\or% 12pt
  \newcommand{\miniscule}{\@setfontsize\miniscule{5}{6}}% \tiny: 6/7
\fi
\makeatother

\DeclareRobustCommand{\optbar}[1]{\shortstack{{\miniscule (\rule[.5ex]{1.25em}{.18mm})}
  \\ [-.7ex] $#1$}}

\def\mumu       {{\ensuremath{\Pmu^+\Pmu^-}}\xspace}

\def\dquark    {{\ensuremath{\Pd}}\xspace}

\def\squark    {{\ensuremath{\Ps}}\xspace}
\def\squarkbar {{\ensuremath{\overline \squark}}\xspace}

\def\cquark    {{\ensuremath{\Pc}}\xspace}

\def\bquark    {{\ensuremath{\Pb}}\xspace}
\def\bquarkbar {{\ensuremath{\overline \bquark}}\xspace}

\def\pion   {{\ensuremath{\Ppi}}\xspace}

\def\pip    {{\ensuremath{\pion^+}}\xspace}
\def\pim    {{\ensuremath{\pion^-}}\xspace}
\def\pipi  {\ensuremath{\pion^+\pion^-}\xspace}

\def\kaon    {{\ensuremath{\PK}}\xspace}
  \def\Kbar    {{\kern 0.2em\overline{\kern -0.2em \PK}{}}\xspace}

\def\KorKbar    {\kern 0.18em\optbar{\kern -0.18em K}{}\xspace}
\def\Kz      {{\ensuremath{\kaon^0}}\xspace}

\def\Kp      {{\ensuremath{\kaon^+}}\xspace}
\def\Km      {{\ensuremath{\kaon^-}}\xspace}

\def\KS      {{\ensuremath{\kaon^0_{\rm\scriptscriptstyle S}}}\xspace}

\def\Kstarz  {{\ensuremath{\kaon^{*}(892)^0}}\xspace}

  \def\Dbar    {{\kern 0.2em\overline{\kern -0.2em \PD}{}}\xspace}

\def\DorDbar    {\kern 0.18em\optbar{\kern -0.18em D}{}\xspace}

\def\B       {{\ensuremath{\PB}}\xspace}
\def\Bbar    {{\ensuremath{\kern 0.18em\overline{\kern -0.18em \PB}{}}}\xspace}
\def\Bb      {{\ensuremath{\Bbar}}\xspace}
\def\BorBbar    {\kern 0.18em\optbar{\kern -0.18em B}{}\xspace}
\def\Bz      {{\ensuremath{\B^0}}\xspace}

\def\Bu      {{\ensuremath{\B^+}}\xspace}
\def\Bub     {{\ensuremath{\B^-}}\xspace}
\def\Bp      {{\ensuremath{\Bu}}\xspace}

\def\Bd      {{\ensuremath{\B^0}}\xspace}
\def\Bs      {{\ensuremath{\B^0_\squark}}\xspace}

\def\Bsb     {{\ensuremath{\Bbar{}^0_\squark}}\xspace}
\def\Bdb     {{\ensuremath{\Bbar{}^0}}\xspace}

\def\jpsi     {{\ensuremath{{\PJ\mskip -3mu/\mskip -2mu\Ppsi\mskip 2mu}}}\xspace}

  \def\Y#1S{\ensuremath{\PUpsilon{(#1S)}}\xspace}% no space before {...}!

\def\FourS {{\Y4S}}

\def\proton      {{\ensuremath{\Pp}}\xspace}

\def\Lbar        {{\ensuremath{\kern 0.1em\overline{\kern -0.1em\PLambda}}}\xspace}
\def\LorLbar    {\kern 0.18em\optbar{\kern -0.18em \PLambda}{}\xspace}

\def\BF         {{\ensuremath{\cal B}}\xspace}

\def\BR         {\BF}
\newcommand{\decay}[2]{\ensuremath{#1\!\to #2}\xspace}         % {\Pa}{\Pb \Pc}

\def\to                 {\ensuremath{\rightarrow}\xspace}

\def\CP                {{\ensuremath{C\!P}}\xspace}

\newcommand{\dms}{{\ensuremath{\Delta m_{\squark}}}\xspace}

\newcommand{\DG}{{\ensuremath{\Delta\Gamma}}\xspace}
\newcommand{\DGs}{{\ensuremath{\Delta\Gamma_{\squark}}}\xspace}
\newcommand{\DGd}{{\ensuremath{\Delta\Gamma_{\dquark}}}\xspace}

\newcommand{\GL}{{\ensuremath{\Gamma_{\rm L}}}\xspace}
\newcommand{\GH}{{\ensuremath{\Gamma_{\rm H}}}\xspace}

\newcommand{\phid}{{\ensuremath{\phi_{\dquark}}}\xspace}

\newcommand{\phis}{{\ensuremath{\phi_{\squark}}}\xspace}

\newcommand{\mistag}{\ensuremath{\omega}\xspace}

\newcommand{\etag}{{\ensuremath{\varepsilon_{\rm tag}}}\xspace}

\newcommand{\effeff}{\ensuremath{\varepsilon_{\rm eff}}\xspace}

\def\AT#1     {\ensuremath{A_{\mathrm{T}}^{#1}}\xspace}           % 2

\def\C#1      {\ensuremath{\mathcal{C}_{#1}}\xspace}                       % 9
\def\Cp#1     {\ensuremath{\mathcal{C}_{#1}^{'}}\xspace}                    % 7
\def\Ceff#1   {\ensuremath{\mathcal{C}_{#1}^{\mathrm{(eff)}}}\xspace}        % 9  
\def\Cpeff#1  {\ensuremath{\mathcal{C}_{#1}^{'\mathrm{(eff)}}}\xspace}       % 7
\def\Ope#1    {\ensuremath{\mathcal{O}_{#1}}\xspace}                       % 2
\def\Opep#1   {\ensuremath{\mathcal{O}_{#1}^{'}}\xspace}                    % 7

\newcommand{\tev}{\ifthenelse{\boolean{inbibliography}}{\ensuremath{~T\kern -0.05em eV}\xspace}{\ensuremath{\mathrm{\,Te\kern -0.1em V}}}\xspace}
\newcommand{\gev}{\ensuremath{\mathrm{\,Ge\kern -0.1em V}}\xspace}
\newcommand{\mev}{\ensuremath{\mathrm{\,Me\kern -0.1em V}}\xspace}
\newcommand{\kev}{\ensuremath{\mathrm{\,ke\kern -0.1em V}}\xspace}
\newcommand{\ev}{\ensuremath{\mathrm{\,e\kern -0.1em V}}\xspace}
\newcommand{\gevc}{\ensuremath{{\mathrm{\,Ge\kern -0.1em V\!/}c}}\xspace}
\newcommand{\mevc}{\ensuremath{{\mathrm{\,Me\kern -0.1em V\!/}c}}\xspace}
\newcommand{\gevcc}{\ensuremath{{\mathrm{\,Ge\kern -0.1em V\!/}c^2}}\xspace}
\newcommand{\gevgevcccc}{\ensuremath{{\mathrm{\,Ge\kern -0.1em V^2\!/}c^4}}\xspace}
\newcommand{\mevcc}{\ensuremath{{\mathrm{\,Me\kern -0.1em V\!/}c^2}}\xspace}

\def\mum  {\ensuremath{{\,\upmu\rm m}}\xspace}

\def\invfb   {\ensuremath{\mbox{\,fb}^{-1}}\xspace}

\def\ps   {\ensuremath{{\rm \,ps}}\xspace}

\def\invps{\ensuremath{{\rm \,ps^{-1}}}\xspace}

\def\gsim{{~\raise.15em\hbox{$>$}\kern-.85em
          \lower.35em\hbox{$\sim$}~}\xspace}
\def\lsim{{~\raise.15em\hbox{$<$}\kern-.85em
          \lower.35em\hbox{$\sim$}~}\xspace}

\newcommand{\Real}{\ensuremath{\mathcal{R}e}\xspace}
\newcommand{\Imag}{\ensuremath{\mathcal{I}m}\xspace}

\def\sPlot{\mbox{\em sPlot}\xspace}
\def\ptot       {\mbox{$p$}\xspace}
\def\pt         {\mbox{$p_{\rm T}$}\xspace}

\def\evtgen     {\mbox{\textsc{EvtGen}}\xspace}

\def\geant      {\mbox{\textsc{Geant4}}\xspace}

\def\photos     {\mbox{\textsc{Photos}}\xspace}

\def\pythia     {\mbox{\textsc{Pythia}}\xspace}

\def\tell1  {TELL1\xspace}
\def\ukl1   {UKL1\xspace}

\usepackage{cite} % Allows for ranges in citations
\usepackage{mciteplus}

\newcommand{\BRBsBdVal}{0.0431}
\newcommand{\BRBsVal}{1.93}
\newcommand{\ADGVal}{\phantom{-}0.49}
\newcommand{\cdirVal}{-0.28}
\newcommand{\smixVal}{-0.08}
\newcommand{\cdirBdVal}{-0.028}
\newcommand{\smixBdVal}{0.719}

\newcommand{\BRBsBdStat}{\pm 0.0017\:(\text{stat})}
\newcommand{\BRBsStat}{\pm 0.08\:(\text{stat})}
\newcommand{\ADGStat}{\pm\:^{0.77}_{0.65}\:\:(\text{stat})}
\newcommand{\cdirStat}{\pm 0.41\:(\text{stat})}
\newcommand{\smixStat}{\pm 0.40\:(\text{stat})}
\newcommand{\cdirBdStat}{\pm 0.034\:(\text{stat})}
\newcommand{\smixBdStat}{\pm 0.034\:(\text{stat})}

\newcommand{\BRBsBdSyst}{\pm 0.0012\:(\text{syst})}
\newcommand{\BRBsSyst}{\pm 0.05\:(\text{syst})}
\newcommand{\ADGSyst}{\pm 0.06\:(\text{syst})}
\newcommand{\cdirSyst}{\pm 0.08\:(\text{syst})}
\newcommand{\smixSyst}{\pm 0.08\:(\text{syst})}
\newcommand{\BRBsBdfds}{\pm 0.0025\:(f_s/f_d)}
\newcommand{\BRBsfds}{\pm 0.11\:(f_s/f_d)}
\newcommand{\BRBsPDG}{\pm 0.07\:(\BR(\Bd\to\jpsi{}\Kz)}

\newcommand{\rawLL}{0.01104 \pm 0.00072\:(\text{stat}) \pm 0.00020\:(\text{syst})}
\newcommand{\rawDD}{0.01170 \pm 0.00059\:(\text{stat}) \pm 0.00019\:(\text{syst})}
\newcommand{\SelLL}{0.972 \pm 0.029}
\newcommand{\SelDD}{0.987 \pm 0.040}

\newcommand{\BdToJPsiKS}{\decay{\Bd}{\jpsi\KS}}
\newcommand{\BsToJPsiKS}{\decay{\Bs}{\jpsi\KS}}

\newcommand{\smix}{\ensuremath{S_{\text{mix}}}\xspace}
\newcommand{\cdir}{\ensuremath{C_{\text{dir}}}\xspace}
\newcommand{\ADG }{\ensuremath{\mathcal{A}_{\Delta\Gamma}}\xspace}
\newcommand{\mass}{\ensuremath{m_{\jpsi\KS}}\xspace}

\begin{document}

%%%%%%%%%%%%%%%%%%%%%%%%%
%%%%% Title     %%%%%%%%%
%%%%%%%%%%%%%%%%%%%%%%%%%
\renewcommand{\thefootnote}{\fnsymbol{footnote}}
\setcounter{footnote}{1}

% %%%%%%% CHOOSE TITLE PAGE--------
%!TEX root=main.tex

\def\thetitle{Measurement of the time-dependent \CP asymmetries in $\BsToJPsiKS$}

%%%%%%%%%%%%%%%%%%%%%%%%%
%%%%%  TITLE PAGE  %%%%%%
%%%%%%%%%%%%%%%%%%%%%%%%%
\begin{titlepage}
\pagenumbering{roman}

% Header ---------------------------------------------------
\vspace*{-1.5cm}
\centerline{\large EUROPEAN ORGANIZATION FOR NUCLEAR RESEARCH (CERN)}
\vspace*{1.5cm}
\hspace*{-0.5cm}
\begin{tabular*}{\linewidth}{lc@{\extracolsep{\fill}}r}
\ifthenelse{\boolean{pdflatex}}% Logo format choice
{\vspace*{-2.7cm}\mbox{\!\!\!\includegraphics[width=.14\textwidth]{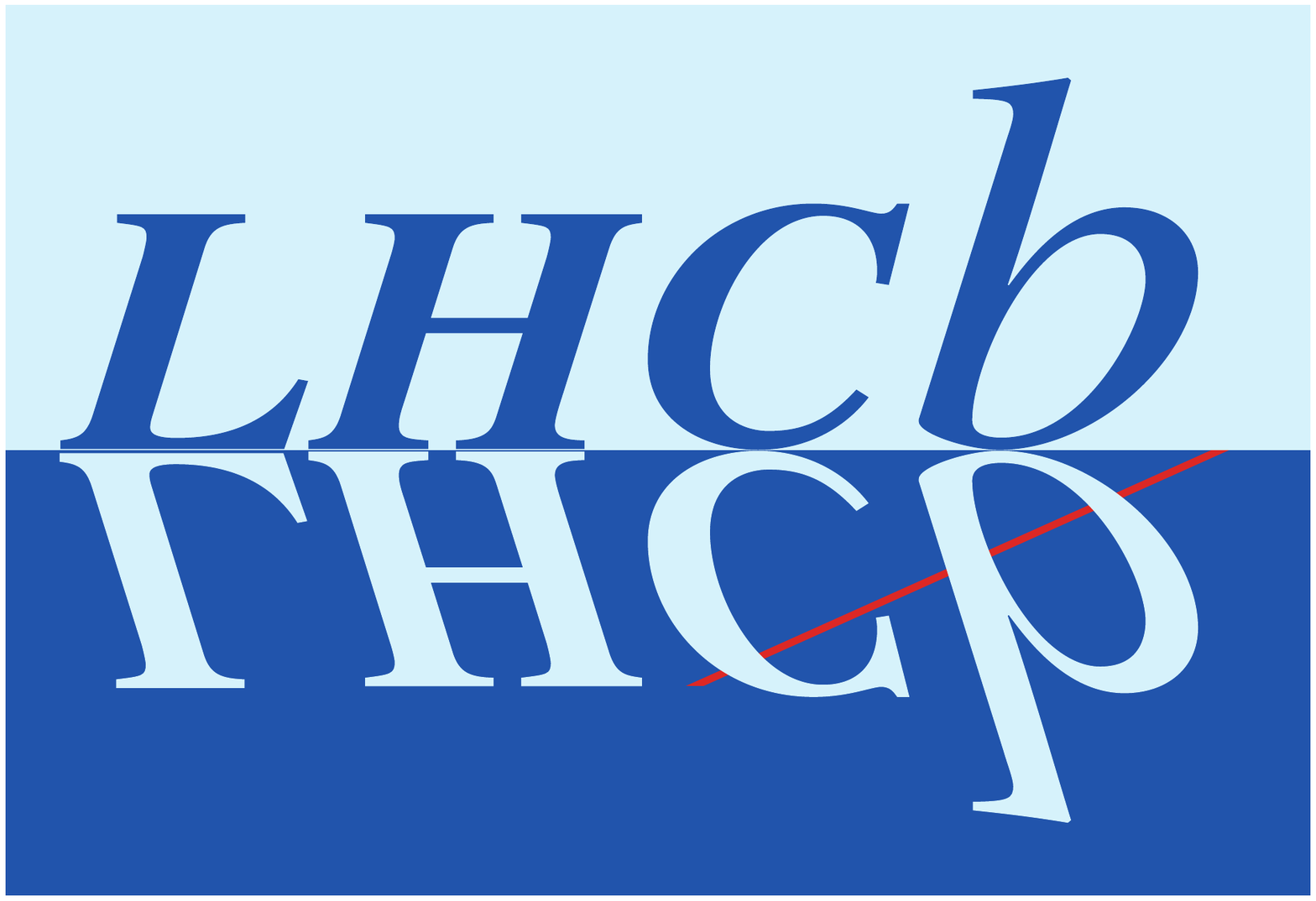}} & &}%
{\vspace*{-1.2cm}\mbox{\!\!\!\includegraphics[width=.12\textwidth]{lhcb-logo.eps}} & &}%
\\
 & & CERN-PH-EP-2015-064 \\  % ID 
 & & LHCb-PAPER-2015-005 \\  % ID 
 & & March 24, 2015 \\ % Date
 & & \\
\end{tabular*}

\vspace*{1.5cm}

% Title --------------------------------------------------
{\bf\boldmath\huge
\begin{center}
  \thetitle
\end{center}
}

\vspace*{1.0cm}

% Authors -------------------------------------------------
\begin{center}
The LHCb collaboration\footnote{Authors are listed at the end of this paper.}
\end{center}

\vspace{\fill}

% Abstract -----------------------------------------------
\begin{abstract}

The first measurement of decay-time-dependent \CP{} asymmetries in the decay \mbox{\BsToJPsiKS} and an updated measurement of the ratio of branching fractions \mbox{$\BR(\BsToJPsiKS)/\BR(\BdToJPsiKS)$} are presented.
The results are obtained using data corresponding to an integrated luminosity of 3.0\:\invfb of proton--proton collisions recorded with the LHCb detector at centre-of-mass energies of 7 and 8 \tev.
The results on the \CP asymmetries are
\begin{align*}
\ADG\left(\BsToJPsiKS\right)  & = \ADGVal \ADGStat \ADGSyst\:,\\
\cdir\left(\BsToJPsiKS\right)  & = \cdirVal \cdirStat \cdirSyst\:,\\
\smix\left(\BsToJPsiKS\right)  & = \smixVal \smixStat \smixSyst\:.
\end{align*}
The ratio \mbox{$\BR(\BsToJPsiKS)/\BR(\BdToJPsiKS)$} is measured to be
\begin{equation*}
\BRBsBdVal \BRBsBdStat \BRBsBdSyst \BRBsBdfds\:,
\end{equation*}
where the last uncertainty is due to the knowledge of the $\Bs$ and $\Bd$ production fractions.
\end{abstract}

\vspace*{1.5cm}

\begin{center}
Published in JHEP 06 (2015) 131
\end{center}

\vspace{\fill}

{\footnotesize 
\centerline{\copyright~CERN on behalf of the \lhcb collaboration, licence \href{http://creativecommons.org/licenses/by/4.0/}{CC-BY-4.0}.}}
\vspace*{2mm}

\end{titlepage}

%%%%%%%%%%%%%%%%%%%%%%%%%%%%%%%%
%%%%%  EOD OF TITLE PAGE  %%%%%%
%%%%%%%%%%%%%%%%%%%%%%%%%%%%%%%%

%  empty page follows the title page ----
\newpage
\setcounter{page}{2}
\mbox{~}

\cleardoublepage

% %%%%%%%%%%%%% ---------

\renewcommand{\thefootnote}{\arabic{footnote}}
\setcounter{footnote}{0}

%%%%%%%%%%%%%%%%%%%%%%%%%
%%%%% Main text %%%%%%%%%
%%%%%%%%%%%%%%%%%%%%%%%%%

\pagestyle{plain} % restore page numbers for the main text
\setcounter{page}{1}
\pagenumbering{arabic}

%% Uncomment during review phase. 
%% Comment before a final submission.
%\linenumbers

%!TEX root=main.tex

%%%%%%%%%%%%%%%%%%%%%%%%%%%%%%%%%%%%%%%%%%%%%%%%%%%%%%%%%%
\section{Introduction}\label{Sec:Introduction}
%%%%%%%%%%%%%%%%%%%%%%%%%%%%%%%%%%%%%%%%%%%%%%%%%%%%%%%%%%

\begin{figure}[t]
\includegraphics[width=0.49\textwidth]{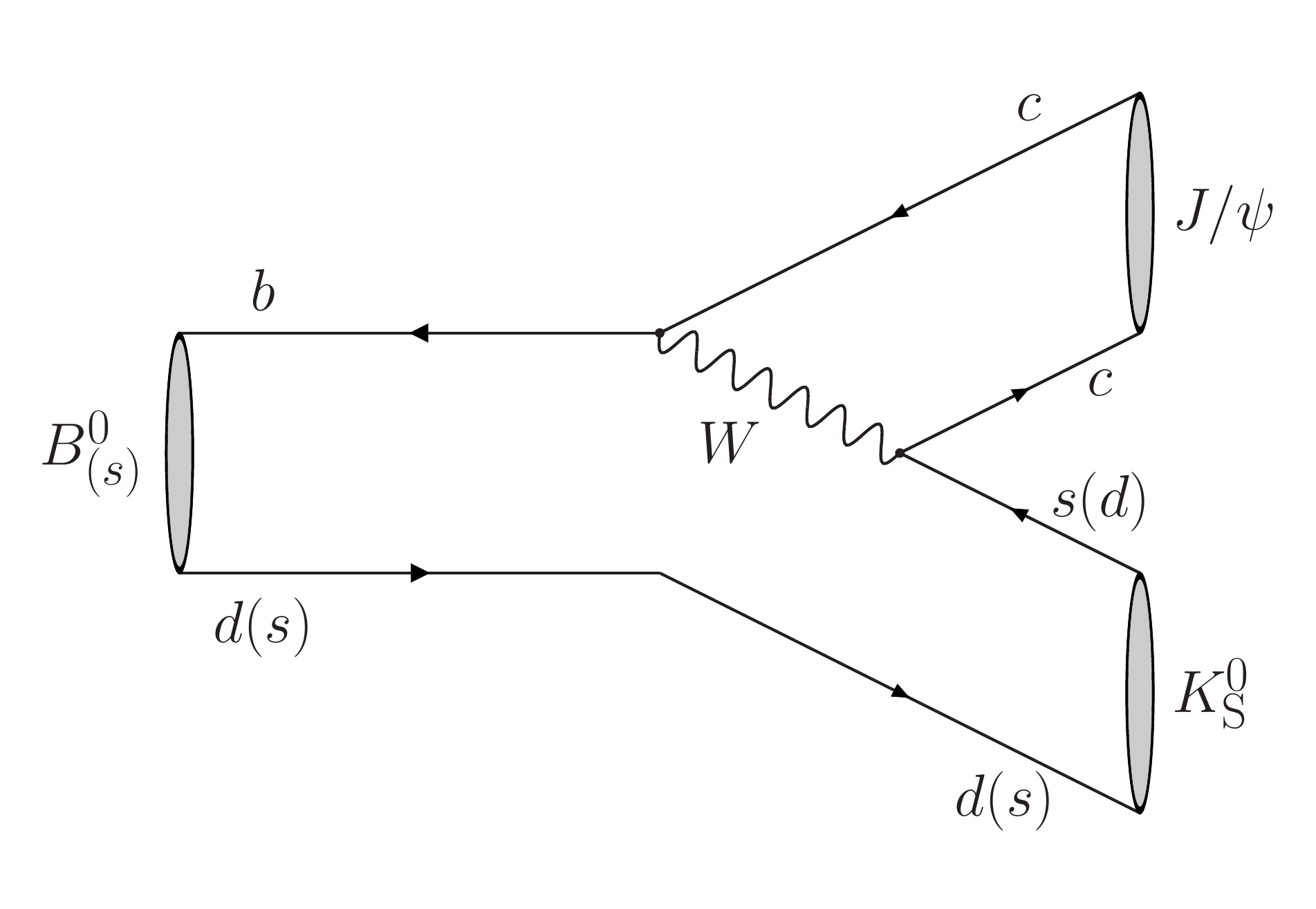}
\hfill
\includegraphics[width=0.49\textwidth]{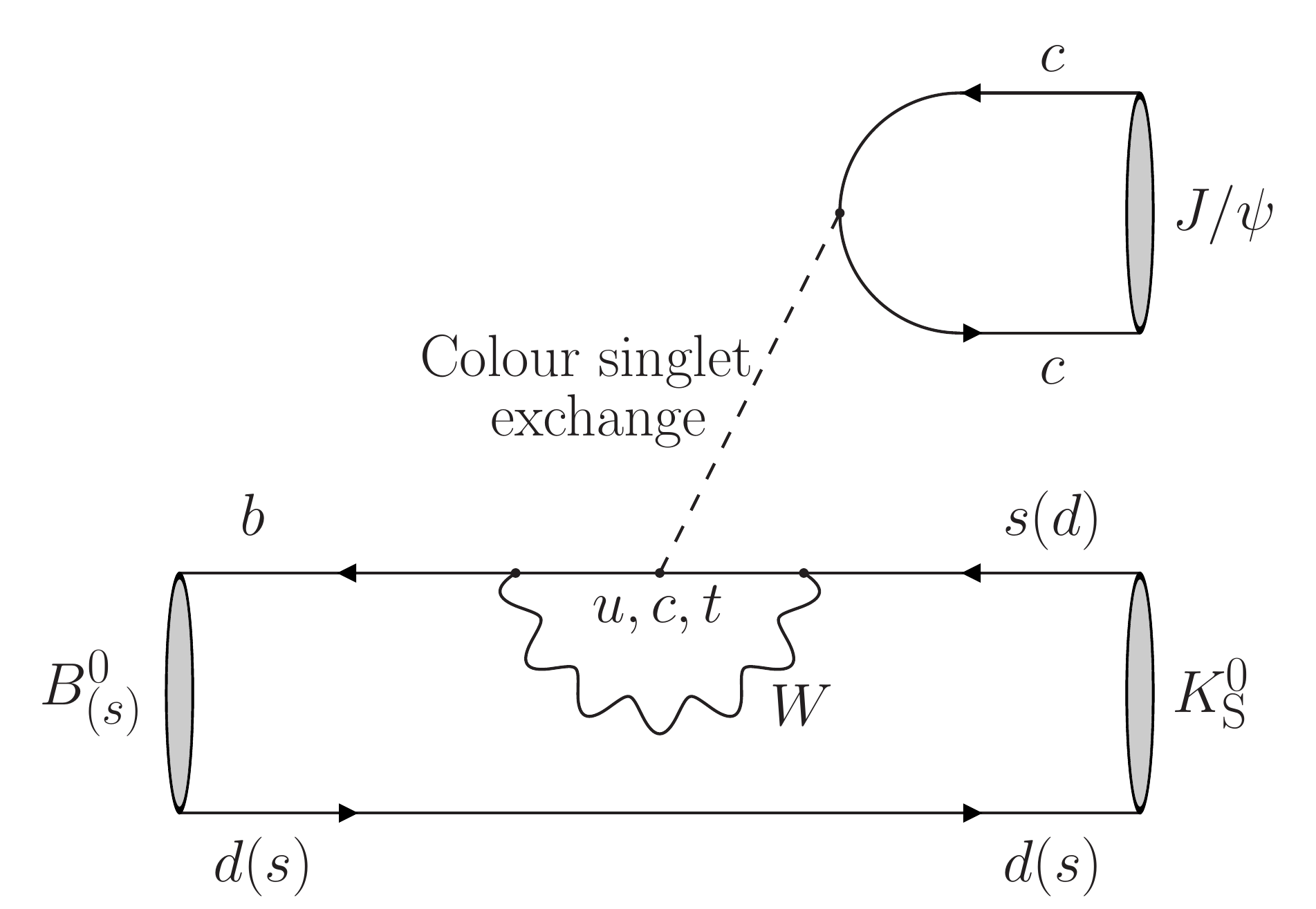}
\caption{Decay topologies contributing to the $\decay{B_{(s)}^0}{\jpsi{}\KS}$ channel:
(left) tree diagram and (right) penguin diagram.}
\label{Fig:Feynman}
\end{figure}

%%% Bd->JpsiKs
In decays of neutral \B mesons (where \B stands for a \Bd or \Bs meson) to a final state accessible to both $\B$ and $\Bbar$, the interference between the direct decay and the decay via oscillation leads to decay-time-dependent \CP{} violation.
Measurements of time-dependent \CP asymmetries provide valuable tests of the flavour sector of the Standard Model (SM) and offer opportunities to search for signs of non-SM physics.
A measurement of this asymmetry in the $\BdToJPsiKS$ decay mode allows for a determination of the effective \CP phase \cite{Faller:2008zc, DeBruyn:2010hh, DeBruyn:2014oga}
\begin{equation}
\phi_d^{\text{eff}}(\BdToJPsiKS) \equiv \phid + \Delta\phid\:,
\end{equation}
where \phid is the relative phase of the $\Bd$--$\Bdb$ mixing amplitude and the tree-level decay process, and $\Delta\phid$ is a shift induced by the so-called penguin topologies, which are illustrated in Fig.~\ref{Fig:Feynman}.
In the Standard Model, \phid is equal to $2\beta$ \cite{Bigi:1981qs}, where $\beta \equiv \arg(-V_{cd}^{\phantom{*}}V_{cb}^*/V_{td}^{\phantom{*}}V_{tb}^*)$ is one of the angles of the unitarity triangle in the Cabibbo-Kobayashi-Maskawa (CKM) quark mixing matrix \cite{Kobayashi:1973fv,*Cabibbo:1963yz}.
The latest average of the Belle and BaBar measurements reads $\sin\phi_d^{\text{eff}} = 0.665 \pm 0.020$ \cite{HFAG}, while the recently updated analysis from LHCb reports $\sin\phi_d^{\text{eff}} = 0.729 \pm 0.035 \text{(stat)} \pm 0.022 \text{(syst)}$ \cite{sin2beta}.

%%% Introducing Penguins
Forthcoming data from the LHC and KEK $e^+e^-$ super $B$ factory will lead to an unprecedented precision on the phase $\phi_d^{\text{eff}}$.
To translate this into an equally precise determination of the CKM phase $\beta$, it is essential to take into account the doubly Cabibbo-suppressed contributions from the penguin topologies, which lead to a value for $\Delta\phid$ that might be as large as $\mathcal{O}(1^{\circ})$ \cite{Faller:2008zc, DeBruyn:2014oga}.
By relying on approximate flavour symmetries, information on $\Delta\phid$ can be obtained from measurements of \CP asymmetries in decays where the penguin topologies are enhanced.
The $\BsToJPsiKS$ mode is the most promising candidate for this task \cite{Fleischer:1999nz, DeBruyn:2010hh, DeBruyn:2014oga}.

%%% CP Observables
Assuming no \CP violation in mixing \cite{HFAG}, the time-dependent \CP asymmetry in \mbox{\BsToJPsiKS} takes the form
\begin{align}
a_{\CP}(t)
& \equiv \frac{\Gamma(\decay{\Bsb(t)}{\jpsi\KS}) - \Gamma(\decay{\Bs(t)}{\jpsi\KS})}{\Gamma(\decay{\Bsb(t)}{\jpsi\KS}) + \Gamma(\decay{\Bs(t)}{\jpsi\KS})}\:,\\
& = \frac{\smix\sin\left(\dms\, t\right) - \cdir\cos\left(\dms\, t\right)}{\cosh\left(\DGs \,t/2\right) + \ADG\sinh\left(\DGs \,t/2\right)}\:,\label{Eq:aCP}
\end{align}
where $\Gamma(\decay{\Bs(t)}{\jpsi\KS})$ represents the time-dependent decay rate of the \Bs meson into the $\jpsi\KS$ final state, and $\dms\equiv m_{\rm H} - m_{\rm L}$ and $\DGs\equiv\GL-\GH$ are, respectively, the mass and decay width difference between the heavy and light eigenstates of the $\Bs$ meson system.
The $\BsToJPsiKS$ \CP observables are defined through the parameter
\begin{equation}
\lambda_{\jpsi\KS} \equiv -e^{i\phis}\:\frac{A(\decay{\Bsb}{\jpsi\KS})}{A(\BsToJPsiKS)}
\end{equation}
in terms of the complex phase $\phis$ associated with the \Bs--\Bsb mixing process and the ratio of time-independent transition amplitudes as
\begin{equation}\label{Eq:CPobs}
\ADG \equiv -\frac{2\,\Real[\lambda_{\jpsi\KS}]}{1+|\lambda_{\jpsi\KS}|^2}\:,\qquad
\cdir \equiv \frac{1-|\lambda_{\jpsi\KS}|^2}{1+|\lambda_{\jpsi\KS}|^2}\:,\qquad
\smix \equiv \frac{2\,\Imag[\lambda_{\jpsi\KS}]}{1+|\lambda_{\jpsi\KS}|^2}\:,
\end{equation}
where $\cdir$ and $\smix$ represent direct and mixing-induced \CP violation, respectively.
In the Standard Model $\phi_s^{\text{SM}} \equiv 2\arg(-V_{ts}^{\phantom{*}}V_{tb}^*)$.
A recent analysis \cite{DeBruyn:2014oga} predicts
\begin{eqnarray}\label{Eq:CPprediction}
\ADG \left(\BsToJPsiKS\right)  & = & 0.957 \pm 0.061\:,\nonumber\\
\cdir\left(\BsToJPsiKS\right) & = & 0.003\pm0.021\:,\\
\smix\left(\BsToJPsiKS\right) & = & 0.29\phantom{0} \pm 0.20\phantom{0}\:.\nonumber
\end{eqnarray}
Similar expression for Eqs.~\eqref{Eq:aCP} and \eqref{Eq:CPobs} are obtained for the $\BdToJPsiKS$ decay by replacing $s\leftrightarrow d$.
The observable $\ADG$ is not applicable in the measurement of $\BdToJPsiKS$ because it is assumed that $\DGd=0$ \cite{HFAG}.

%%%%%%%%%%%%%%%%%%%%%%%%%%%%%%%%%%%%%%%%%%%%%%%%%%%%%%%%%%
%%%%%%%%%%%%%%%%%%%%%%%%%%%%%%%%%%%%%%%%%%%%%%%%%%%%%%%%%%

This paper presents the first measurement of the time-dependent \CP asymmetries in \BsToJPsiKS decays, as well as an updated measurement of the ratio of time-integrated branching fractions \mbox{$\BR(\BsToJPsiKS)/\BR(\BdToJPsiKS)$}.
This ratio was first measured by the CDF collaboration \cite{Aaltonen:2011sy}, while the previously most precise measurement was reported by LHCb in Ref.~\cite{LHCb-PAPER-2013-015}.
The analysis is performed with a data sample corresponding to an integrated luminosity of $3.0\:\invfb$ of proton--proton ($pp$) collisions, recorded by the LHCb experiment at centre-of-mass energies of $7\:\tev$ and $8\:\tev$ in 2011 and 2012, respectively.

The analysis proceeds in two steps.
The first step, described in detail in Sec.~\ref{Sec:Selection}, consists of a multivariate selection of $\decay{\B}{\jpsi\KS}$ candidates.
In the second step a maximum likelihood fit is performed to the selected data.
The fit model includes a prominent $\BdToJPsiKS$ component, which is used to improve the modelling of the $\BsToJPsiKS$ signal. In addition, the measurement of \CP asymmetries associated with $\BdToJPsiKS$ decays offers a validation of the likelihood method's implementation.
However, the stringent event selection necessary to isolate the $\BsToJPsiKS$ candidates limits the precision on these two \CP observables.
Dedicated and more precise measurements of the $\BdToJPsiKS$ \CP observables are therefore the subject of a separate publication \cite{sin2beta}.

For a time-dependent measurement of \CP violation it is essential to determine the initial flavour of the $\B$ candidate, i.e.\ whether it contained a $\bquark$ or a $\bquarkbar$ quark at production.
The method to achieve this is called \emph{flavour tagging}, and is discussed in Sec.~\ref{Sec:Tagging}.
The tagging information is combined with a description of the \B mass and decay time distributions when performing the maximum likelihood fit, which is described in Sec.~\ref{Sec:PDF}.
The three \CP observables describing the $\BsToJPsiKS$ decays and two \CP observables describing the $\BdToJPsiKS$ decays are obtained directly from the fit.
The ratio of branching fractions \cite{DeBruyn:2012wj} is derived from the ratio $R$ of fitted $\BsToJPsiKS$ to $\BdToJPsiKS$ event yields as
\begin{equation}\label{Eq:BR_Def}
\frac{\BR(\BsToJPsiKS)}{\BR(\BdToJPsiKS)} = R \times f_{\text{sel}} \times \frac{f_d}{f_s}\:,
\end{equation}
where $f_{\text{sel}}$ is a correction factor for differences in selection efficiency between $\BdToJPsiKS$ and $\BsToJPsiKS$ decays, and $f_s/f_d = 0.259 \pm 0.015$ \cite{LHCb-PAPER-2012-037, *LHCb-CONF-2013-011} is the ratio of \Bs to \Bd meson hadronisation fractions.
The study of systematic effects on the ratio $R$ and the \CP observables is presented in Sec.~\ref{Sec:Syst}.
The main results for the branching ratio measurement are reported in Sec.~\ref{Sec:BR} and those for the \CP observables in Sec.~\ref{Sec:Conclusion}.

%!TEX root=main.tex

%%%%%%%%%%%%%%%%%%%%%%%%%%%%%%%%%%%%%%%%%%%%%%%%%%%%%%%%%%
\section{Detector and simulation} \label{Sec:Detector}
%%%%%%%%%%%%%%%%%%%%%%%%%%%%%%%%%%%%%%%%%%%%%%%%%%%%%%%%%%

The \lhcb detector~\cite{Alves:2008zz,LHCb-DP-2014-002} is a single-arm forward
spectrometer covering the pseudorapidity range $2<\eta <5$,
designed for the study of particles containing \bquark or \cquark
quarks. The detector includes a high-precision tracking system
consisting of a silicon-strip vertex detector surrounding the $pp$
interaction region, a large-area silicon-strip detector located
upstream of a dipole magnet with a bending power of about
$4{\rm\,Tm}$, and three stations of silicon-strip detectors and straw
drift tubes placed downstream of the magnet.
The tracking system provides a measurement of momentum, \ptot, of charged particles with
a relative uncertainty that varies from 0.5\% at low momentum to 1.0\% at 200\gevc.
The minimum distance of a track to a primary vertex, the impact parameter, is measured with a resolution of $(15+29/\pt)\mum$,
where \pt is the component of the momentum transverse to the beam, in \gevc.
Different types of charged hadrons are distinguished using information
from two ring-imaging Cherenkov detectors. 
Photons, electrons and hadrons are identified by a calorimeter system consisting of
scintillating-pad and preshower detectors, an electromagnetic
calorimeter and a hadronic calorimeter. Muons are identified by a
system composed of alternating layers of iron and multiwire
proportional chambers.

In the simulation, $pp$ collisions are generated using
\pythia~\cite{Sjostrand:2006za,*Sjostrand:2007gs} with a specific \lhcb
configuration~\cite{LHCb-PROC-2010-056}.  Decays of hadronic particles
are described by \evtgen~\cite{Lange:2001uf}, in which final-state
radiation is generated using \photos~\cite{Golonka:2005pn}. The
interaction of the generated particles with the detector, and its response,
are implemented using the \geant
toolkit~\cite{Allison:2006ve, *Agostinelli:2002hh} as described in
Ref.~\cite{LHCb-PROC-2011-006}.

%!TEX root=main.tex

%%%%%%%%%%%%%%%%%%%%%%%%%%%%%%%%%%%%%%%%%%%%%%%%%%%%%%%%%%
\section{Event selection}\label{Sec:Selection}
%%%%%%%%%%%%%%%%%%%%%%%%%%%%%%%%%%%%%%%%%%%%%%%%%%%%%%%%%%

Candidate $\decay{\B}{\jpsi{}\KS}$ decays are considered in the $\decay{\jpsi}{\mumu}$ and $\decay{\KS}{\pipi}$ final states. The event selection is based on an initial selection, followed by a two-stage multivariate analysis consisting of artificial neural network (NN) classifiers \cite{Feindt:2004nb}.

%%%%%%%%%%%%%%%%%%%%%%%%%
\subsection{Initial selection}

%%% Trigger
The online event selection is performed by a trigger, which consists of a hardware level, based on information from the calorimeter and muon systems, followed by a software level, which applies a full event reconstruction.
The hardware trigger selects at least one muon with a transverse momentum $\pt>1.48\:(1.76) \gevc$ or two muons with $\sqrt{\pt(\mu_1)\pt(\mu_2)}>1.3\:(1.6)\gevc$ in the 7 (8) \tev $pp$ collisions.
The software trigger consists of two stages.
In the first stage, events are required to have either two oppositely charged muons with combined mass above $2.7\:\gevcc$, or at least one muon or one high-\pt charged particle ($\pt>1.8\:\gevc$) with an impact parameter larger than $100\mum$ with respect to all $pp$ interaction vertices (PVs).
In the second stage of the software trigger the tracks of two or more of the final-state particles are required to form a vertex that is significantly displaced from the PVs, and only events containing $\decay{\jpsi}{\mumu}$ candidates are retained.

%%% Reconstruction & Loose - Jpsi
In the offline selection, $\jpsi$ candidates are selected by requiring two muon tracks to form a good quality vertex and have an invariant mass in the range $[3030,3150]\:\mevcc$.
This interval corresponds to about eight times the $\mumu$ mass resolution at the \jpsi mass and covers part of the $\jpsi$ radiative tail.

%%% KS: long vs downstream
Decays of \decay{\KS}{\pip\pim} are reconstructed in two different categories:
the first involving \KS mesons that decay early enough for the daughter pions to be reconstructed in the vertex detector; and the second containing \KS that decay later such that track segments of the pions cannot be formed in the vertex detector.
These categories are referred to as \emph{long} and \emph{downstream}, respectively.
Long $\KS$ candidates have better mass, momentum and vertex resolution than those in the downstream category.

%%% Reconstruction & Loose - KS
The two pion tracks of the long (downstream) \KS candidates are required to form a good quality vertex and their combined invariant mass must be within $35\:(64)\:\mevcc$ of the known $\KS$ mass \cite{PDG2014}.
To remove contamination from $\decay{\Lambda}{\proton\pim}$ decays, the reconstructed mass of the long (downstream) \KS candidates under the assumption that one of its daughter tracks is a proton is required to be more than $6\:(10)\:\mevcc$ away from the known $\Lambda$ mass \cite{PDG2014}.
The \KS decay vertex is required to be located downstream of the $\jpsi$ decay vertex, i.e.\ it is required to have a positive flight distance.
This removes approximately 50\% of mis-reconstructed $\decay{\Bd}{\jpsi\Kstarz}$ background.
The remaining $\decay{\Bd}{\jpsi\Kstarz}$ background is heavily suppressed by the first stage of the multivariate selection described below.

%%% Reconstruction & Loose - Bd
Candidate \B mesons are selected from combinations of \jpsi{} and \KS candidates with mass $\mass$ in the range $[5180,5520]\:\mevcc$ and a decay time larger than $0.2\:\ps$.
The reconstructed mass and decay time are obtained from a kinematic fit \cite{Hulsbergen:2005pu} that constrains the masses of the \mumu and \pipi pairs to the known \jpsi and \KS masses \cite{PDG2014}, respectively, and constrains the \B candidate to originate from the PV.
A good quality fit is required and the uncertainty on the \B mass estimated by the kinematic fit must not exceed $30\:\mevcc$.
In the case that the event has multiple PVs, a clear separation of the \jpsi decay vertex from any of the other PVs in the event is required, and all combinations of $B$ candidates and PVs that pass the selection are considered.

%%%%%%%%%%%%%%%%%%%%%%%%%
\subsection{Multivariate selection}

%%% B2JpsiKstar Neural Net
The first stage of the multivariate selection focuses on removing the mis-reconstructed $\decay{\Bd}{\jpsi\Kstarz}$ background that survives the requirement on the \KS flight distance.
It only affects the subsample of candidates for which the \KS is reconstructed in the long category.
The NN is trained on simulated $\BdToJPsiKS$ (signal) and $\decay{\Bd}{\jpsi\Kstarz}$ (background) data and only uses information associated with the reconstructed pions and \KS candidate.
This includes decay time, mass, momentum, impact parameter and particle-identification properties.
The requirement on the NN classifier's output is optimised to retain 99\% of the original signal candidates in simulation, with a background rejection on simulated $\decay{\Bd}{\jpsi\Kstarz}$ candidates of 99.55\%.
This results in an estimated number of $18 \pm 2$ $\decay{\Bd}{\jpsi\Kstarz}$ candidates in the long \KS data sample surviving this stage of the selection.
Their yield is further reduced by the second NN classifier, and these candidates are therefore treated as combinatorial background in the remainder of the analysis.
% \newline

%%% sWeights
The second stage of the multivariate selection aims at reducing the combinatorial background to isolate the small $\BsToJPsiKS$ signal.
In contrast to the first NN, it is trained entirely on data, using the $\BdToJPsiKS$ signal as a representative of the signal features of the $\BsToJPsiKS$ decay.
Candidates for the training sample are those populating the mass ranges $[5180,5340]\:\mevcc$ and $[5390,5520]\:\mevcc$, avoiding the \Bs signal region.
The signal and background weights for the training of the second NN are determined using the \sPlot technique \cite{Pivk:2004ty} and obtained by performing an unbinned maximum likelihood fit to the \B mass distribution of the candidates meeting the selection criteria on the first NN classifier's output.
%%% PDF
The fit function is defined as the sum of a \Bd signal component and a combinatorial background where the parametrisation of the individual components matches that of the likelihood method used for the full \CP analysis and is described in more detail in Sec.~\ref{Sec:PDF}.

%%% NeuroBayes - Training
Due to differences in the distributions of the input variables of the NN, as well as different signal-to-background ratios, the second stage of the multivariate selection is performed separately for the \B candidate samples containing long and downstream \KS candidates.
The NN classifiers use information on the candidate's kinematic properties, vertex and track quality, impact parameter, particle identification information from the RICH and muon detectors, as well as global event properties like track and PV multiplicities.
The variables that are used in the second NN are chosen to avoid correlations with the reconstructed \B mass.

\begin{figure}[tp]
\includegraphics[width=0.49\textwidth]{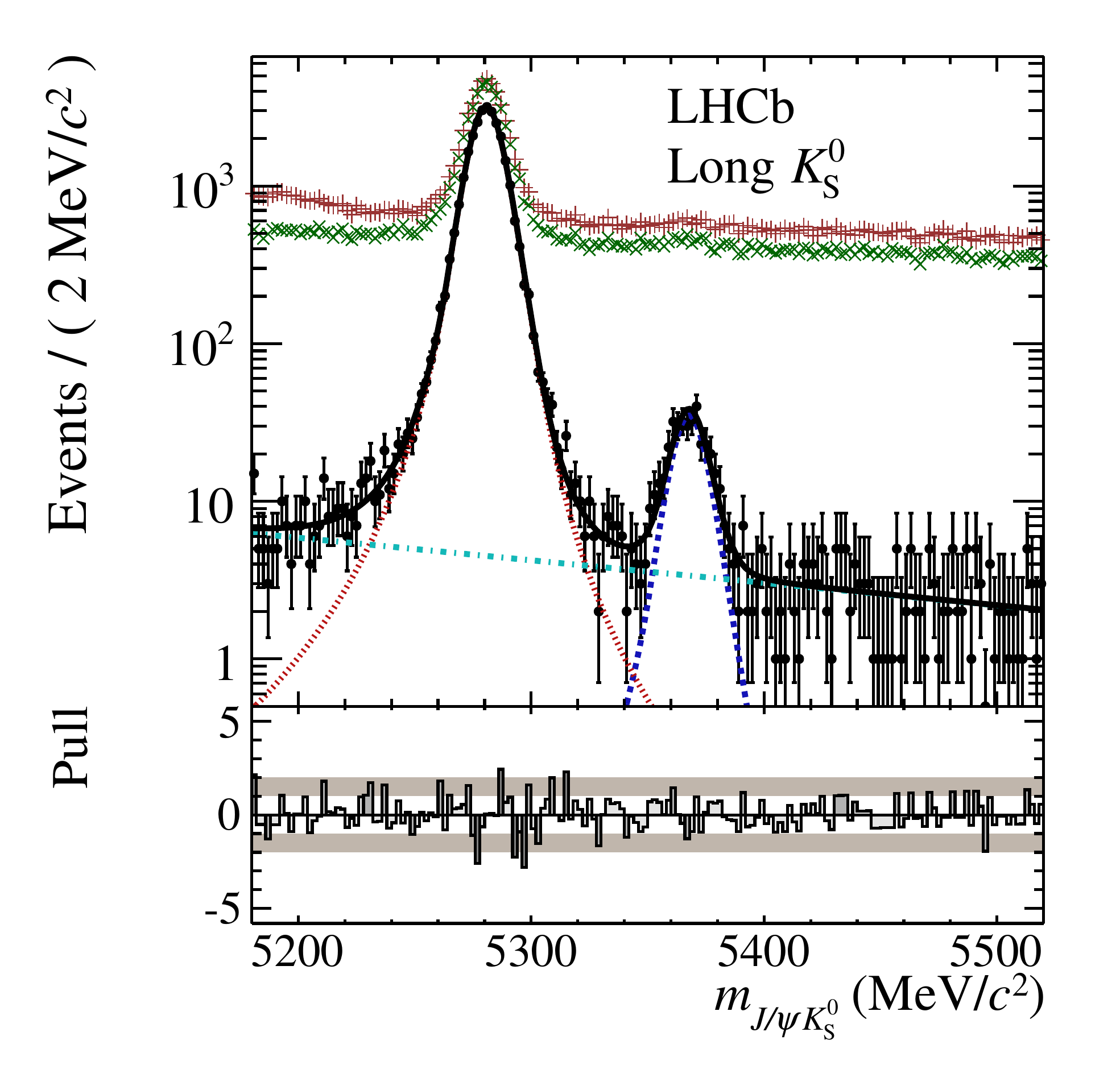}
\hfill
\includegraphics[width=0.49\textwidth]{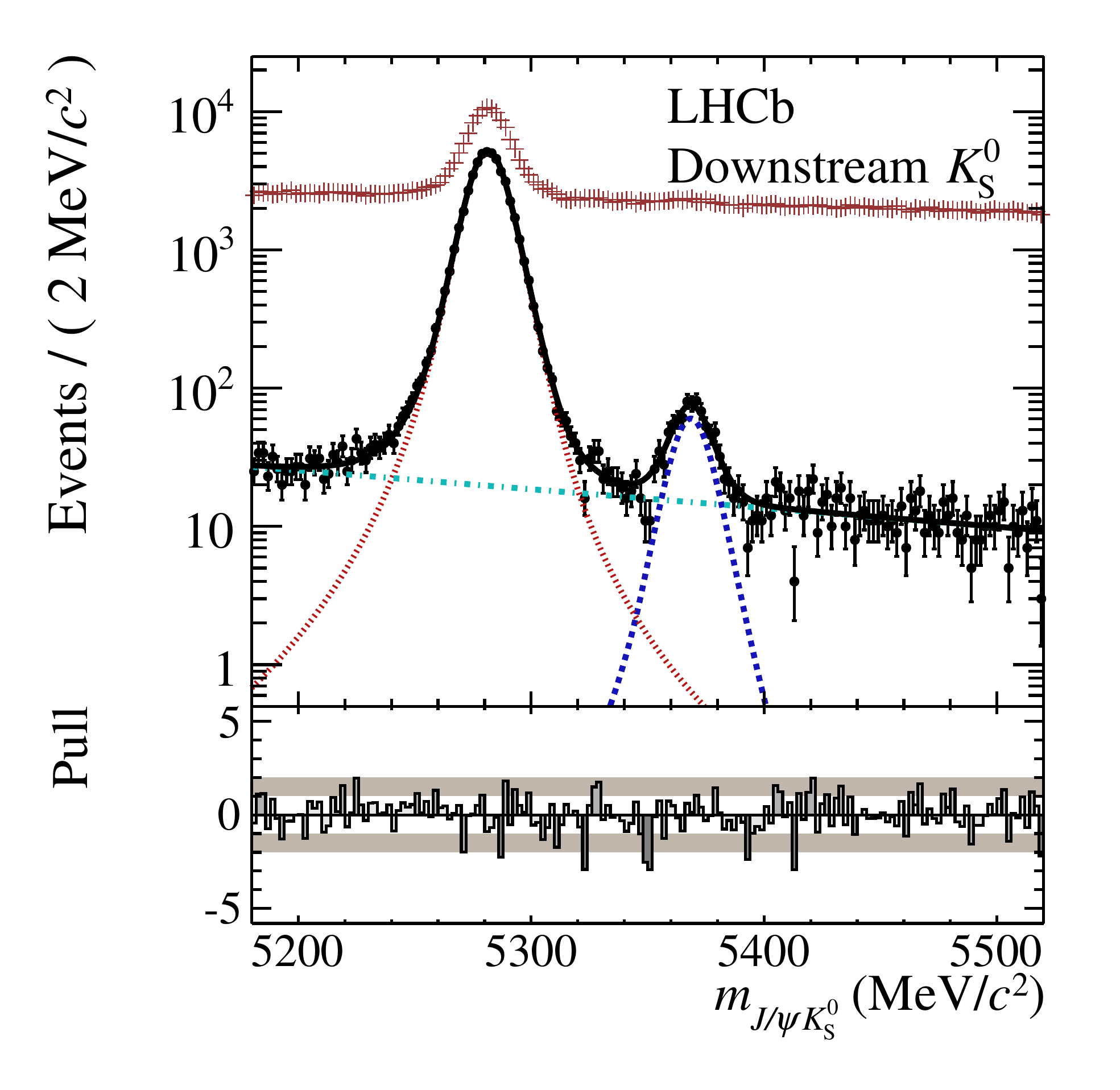}
\caption{Mass distribution of $\B$ candidates at different stages of the event selection for the (left) long \KS and (right) downstream \KS sample.
The data sample after initial selection (red, $+$), after the first neural net (green, $\times$) and after the second neural net (black, $\bullet$) are shown.
Overlaid are projections of the fit described in Sec.~\ref{Sec:PDF}.
Shown components are \BsToJPsiKS (dark blue, dashed), \BdToJPsiKS (red, dotted) and combinatorial background (turquoise, dash-dotted).}
\label{Fig:Mass}
\end{figure}

%%% Optimisation
Final selection requirements on the second stage NN classifier outputs are chosen to optimise the sensitivity to the \Bs signal using $N_\text{S}/\sqrt{N_\text{S}+N_\text{B}}$ as figure of merit, where $N_\text{S}$ and $N_\text{B}$ are respectively the expected number of signal and background events in a $\pm 30\:\mevcc$ mass range around the \Bs peak.
After applying the final requirement on the NN classifier output associated with the long (downstream) \KS sample, the multivariate selection rejects, relative to the initial selection, 99.2\% of the background in both samples while keeping 72.9\% (58.3\%) of the \Bd signal.
The lower selection efficiency on the downstream \KS sample is due to the worse signal-to-background ratio after the initial selection, which requires a more stringent requirement on the NN classifier output.
The resulting $\jpsi{}\KS$ mass distributions are illustrated in Fig.~\ref{Fig:Mass}.

%%%%%%%%%%%%%%%%%%%%%%%%%
%%%%%%%%%%%%%%%%%%%%%%%%%

%%% Multiple Candidates/PVs
After applying the full selection, the long (downstream) \B candidate can still be associated with more than one PV in about 1.5\% (0.6\%) of the events; in this case, one PVs is chosen at random.
Likewise, about $0.24\%$ (0.15\%) of the selected events have multiple candidates sharing one or more tracks; in this case, one candidates is chosen at random.

%!TEX root=main.tex

%%%%%%%%%%%%%%%%%%%%%%%%%%%%%%%%%%%%%%%%%%%%%%%%%%%%%%%%%%
\section{Flavour tagging}\label{Sec:Tagging}
%%%%%%%%%%%%%%%%%%%%%%%%%%%%%%%%%%%%%%%%%%%%%%%%%%%%%%%%%%

%%% Introduction from Bs->DsK paper

At the LHC, $b$ quarks are predominantly produced in $b\bar{b}$ pairs.
When one of the two quarks hadronises to form the $B$ meson decay of interest (``the
signal $B$''), the other $b$ quark hadronises and decays independently.
By exploiting this production mechanism, the signal \B's initial flavour is
identified by means of two classes of flavour-tagging algorithms. The opposite
side (OS) taggers determine the flavour of the non-signal $b$-hadron \cite
{LHCb-PAPER-2011-027} while the same side kaon (SSK) tagger exploits the fact
that the additional $\squarkbar$ ($\squark$) quark produced in the
fragmentation of a $\Bs$ ($\Bsb$) meson often forms a $\Kp$ ($\Km$) meson
\cite {LHCb-CONF-2012-033}.

These algorithms provide tag decisions $q_\text{OS}$ and $q_\text{SSK}$,
which take the value $+1$ ($-1$) in case the signal candidate is tagged as a
$B$ ($\Bb$) meson, and predictions $\eta_\text{OS}$ and $\eta_\text{SSK}$ for the
probability of the tag to be incorrect. The latter is obtained using neural
networks, which in the case of the OS taggers are trained on $\decay{\Bp}{\jpsi\Kp}$
decays, while for the SSK tagger simulated $\decay{\Bs}{D_s^- \pi^+}$ events are
used.

The mistag probability predicted by the tagging algorithms is calibrated in
data to determine the true mistag probability $\omega$, by using control
samples of several flavour-specific $\B$ mesons decays.
This calibration is performed individually for the OS and SSK tagging
algorithms; for the latter, different calibration parameters are used to
describe the \Bz and \Bs mesons. For all events with both an OS and SSK tag
decision, a combined tag decision and mistag probability is derived as
described in Ref.~\cite{LHCb-PAPER-2011-027}.

The figure of merit for the optimisation of a tagging algorithm is the
effective tagging efficiency, $\effeff = \etag (1 - 2 \mistag)^2$ where
$\etag$ is the fraction of candidates with an assigned tag decision. In the
long \KS sample for the \mbox{\BsToJPsiKS} mode, the OS and SSK taggers yield an $\effeff$ of
$(2.93 \pm 0.06)\%$ and $(0.97 \pm 0.12)\%$, respectively, while the sample
with both an OS and SSK tagging response gives an $\effeff$ of $(1.02 \pm
0.10)\%$. In the respective downstream \KS sample, the OS and SSK taggers yield an $\effeff$ of
$(2.74 \pm 0.11)\%$ and $(1.45 \pm 0.15)\%$, respectively, while the sample
with both an OS and SSK tagging response gives an $\effeff$ of $(0.48 \pm
0.04)\%$. The combined $\effeff$ of all three overlapping samples for the
\mbox{\BsToJPsiKS} mode is measured to be $(3.80 \pm 0.18)\%$ and $(4.03
\pm 0.16)$\% in the long and downstream \KS sample, respectively.

% OS, DD:  $(2.74 \pm 0.11)\%$
% OS, LL:  $(2.93 \pm 0.06)\%$
% SSK, DD: $(1.45 \pm 0.15)\%$
% SSK, LL: $(0.97 \pm 0.12)\%$
% OVL, DD: $(0.48 \pm 0.04)\%$
% OVL, LL: $(1.02 \pm 0.10)\%$

% SSK, DD: $(0.098 \pm 0.013)\%$
% SSK, LL: $(0.064 \pm 0.009)\%$

In the $\BdToJPsiKS$ mode, the main contribution is provided by the OS
taggers, where the combined $\effeff$ is measured to be $(2.60 \pm 0.05)\%$
and $(2.63 \pm 0.05)$\% in the long and downstream \KS sample, respectively.
Although the SSK tagging algorithm is specifically designed for \Bs mesons, a
small, but non-vanishing effective tagging efficiency of $(0.064 \pm 0.009)\%$
and $(0.098 \pm 0.013)\%$ in the long and downstream \KS sample, respectively,
is also found for \Bd mesons if the tag decision is reversed. This effect
originates from same-side protons mis-identified as kaons, and kaons from the
decay of \Kstarz mesons produced in correlation with the \Bd. Both tagged
particles have a charge opposite to those of kaons produced in
correlation with the \Bs, and thus require the SSK tag decision to be
inverted. Additionally, mis-identified pions carrying the same charge as the
kaons correlated with the \Bs dilute the effect described above. The SSK
tagging response for \Bd candidates is studied on $\decay{\Bd}{\jpsi\Kstarz}$
candidates using both data and simulated events.

%!TEX root=main.tex

%%%%%%%%%%%%%%%%%%%%%%%%%%%%%%%%%%%%%%%%%%%%%%%%%%%%%%%%%%
\section{Likelihood fit}\label{Sec:PDF}
%%%%%%%%%%%%%%%%%%%%%%%%%%%%%%%%%%%%%%%%%%%%%%%%%%%%%%%%%%

%%% Overview
The $\BsToJPsiKS$ \CP observables are determined from an unbinned maximum likelihood fit.
The data is fitted with a probability density function (PDF) defined as the sum of a \Bd signal component, a \Bs signal component and a combinatorial background.
In total it depends on seven observables. The PDF describes the reconstructed
\B mass $(\mass\in[5180,5520]\:\mevcc)$, the decay time
$(t\in[0.2,15]\:\text{ps})$, and tagging responses $q_\text{OS}$ and
$q_\text{SSK}$. Additionally, it also depends on the per-candidate
decay time uncertainty estimate $\delta_t$ and mistag estimates
$\eta_\text{OS}$ and $\eta_\text{SSK}$.
The long and downstream \KS samples are modelled using separate PDFs but fitted simultaneously.
The parameters common to both PDFs are the two $\BdToJPsiKS$ and three $\BsToJPsiKS$ \CP{} observables, as well as the observables describing the \Bd and \Bs systems that are listed in Table \ref{Tab:Physics_Input}.

%%%%%%%%%%%%%%%%%%%%%%%%%
\begin{table}[t]
\center
\caption{List of the observables describing the \Bd and \Bs systems that are included as Gaussian constraints to the likelihood fit.}
\label{Tab:Physics_Input}
\begin{tabular}{lc@{$\:$}l@{$\:$}r@{$\quad\quad$}lc@{$\:$}l@{$\:$}r}
\toprule
Parameter & \multicolumn{2}{c}{Value} & & Parameter & \multicolumn{2}{c}{Value} &\\
\midrule
  $\Delta m_d$     & $ 0.510 \pm 0.003$ & \invps & \cite{HFAG}
& $\Delta m_s$     & $17.757 \pm 0.021$ & \invps & \cite{HFAG} \\
  $\Delta\Gamma_d$ & $0\phantom{.000\pm 0.000}$   & \invps &
& $\Delta\Gamma_s$ & $\phantom{0}0.081 \pm 0.006$ & \invps & \cite{HFAG} \\
  $\tau_{B_d}$     & $ 1.520 \pm 0.004$ & \ps    & \cite{HFAG}
& $\tau_{B_s}$     & $\phantom{0}1.509 \pm 0.004$ & \ps    & \cite{HFAG} \\
 \bottomrule
\end{tabular}
\end{table}
%%%%%%%%%%%%%%%%%%%%%%%%%

%%%%%%%%%%%%%%%%%%%%%%%%%
\subsection{Mass PDF}

The mass shapes of the $\decay{\B}{\jpsi\KS}$ modes in both data and simulation exhibit non-Gaussian tails on both sides of their signal peaks due to final-state radiation, the detector resolution and its dependence on the momenta of the final-state particles.
Each signal shape is parametrised by a Hypatia function \cite{Santos:2013gra}, whose tail parameters are taken from simulation.
The \Bs component is constrained to have the same shape as the \Bd PDF, but shifted by the \Bs--\Bd mass difference, which is a free variable in the fit.
The mass distribution of the combinatorial background is described by an exponential function.

%%%%%%%%%%%%%%%%%%%%%%%%%
\subsection{Decay time PDF}

The decay time distributions of the two signal components, $\mathcal{T}(t,q_{\text{OS}}, q_{\text{SSK}}|\eta_{\text{OS}},\eta_{\text{SSK}})$,  need to be corrected for experimental effects originating from the detector response and the event selection.
This is done by convolving them with a resolution model, $\mathcal{R}(t|\delta_t)$, and combining the result with an acceptance function, $\mathcal{E}(t)$, to give the experimentally observed decay-time distribution
\begin{equation}
\left(\int\mathcal{T}(\hat{t},q_{\text{OS}}, q_{\text{SSK}}|\eta_{\text{OS}},\eta_{\text{SSK}})\times\mathcal{R}(t-\hat{t}|\delta_t)\:\text{d}\hat{t}\right)\times\mathcal{E}(t)\:.
\end{equation}
%%% Resolution
The resolution model has an individual width for each candidate, described by the per-candidate decay-time uncertainty estimate $\delta_t$ provided by the kinematic fit introduced in Sec.~\ref{Sec:Selection}. A finite resolution reduces the amplitude of the oscillating terms in the decay-time distribution by a factor $\mathcal{D}\equiv \exp\left(-\delta_t^2\Delta m^2/2\right)$ \cite{Moser:1996xf, LHCb-PAPER-2013-002}, and thereby affects the precision of the time-dependent \CP observables.
This effect is larger for the rapid \Bs--\Bsb oscillations than for the \Bd--\Bdb oscillations.
The $\delta_t$ estimates are calibrated using a separate sample of prompt \jpsi{} decays, which are produced directly at the PV and combined with random \KS candidates.
This sample is obtained through the same event selection as described in Sec.~\ref{Sec:Selection}, except for the requirement on the decay time of the \B candidates.
The decay time distribution of the prompt \jpsi{} mesons is modelled by the sum of three Gaussian functions sharing a common mean.
For the long (downstream) \KS sample, this resolution model leads to an average dilution factor of $\langle\mathcal{D}\rangle = 0.73 \pm 0.13$ $(0.72 \pm 0.04)$.

%%% Acceptance
The decay time distribution of the two signal components is affected by acceptance effects due to the decay-time bias induced by the trigger selection, the initial selection requirements and, most importantly, the NN classifier outputs.
The shapes of the $\Bd$ and $\Bs$ acceptances are assumed to be equal and modelled using cubic b-splines \cite{Karbach:2014qba}.
The acceptance function is obtained directly from the data.
The $\BdToJPsiKS$ decay time distribution is described by a single exponential, assuming $\DG_d=0$.
The lifetime of the \Bd, $\tau_{\Bd} = 1.520 \pm 0.004\:\text{ps}$ \cite{HFAG}, is constrained in the fit using a Gaussian function whose mean is fixed to the known lifetime and whose width accounts for the experimental uncertainty.
This allows the acceptance parameters to be directly evaluated in the fit to the data.

%%% Background
The background decay-time distributions are modelled using two exponential functions, describing empirically a short-lived and a long-lived component.

%%%%%%%%%%%%%%%%%%%%%%%%%
\begin{figure}[tp]
\includegraphics[width=0.49\textwidth]{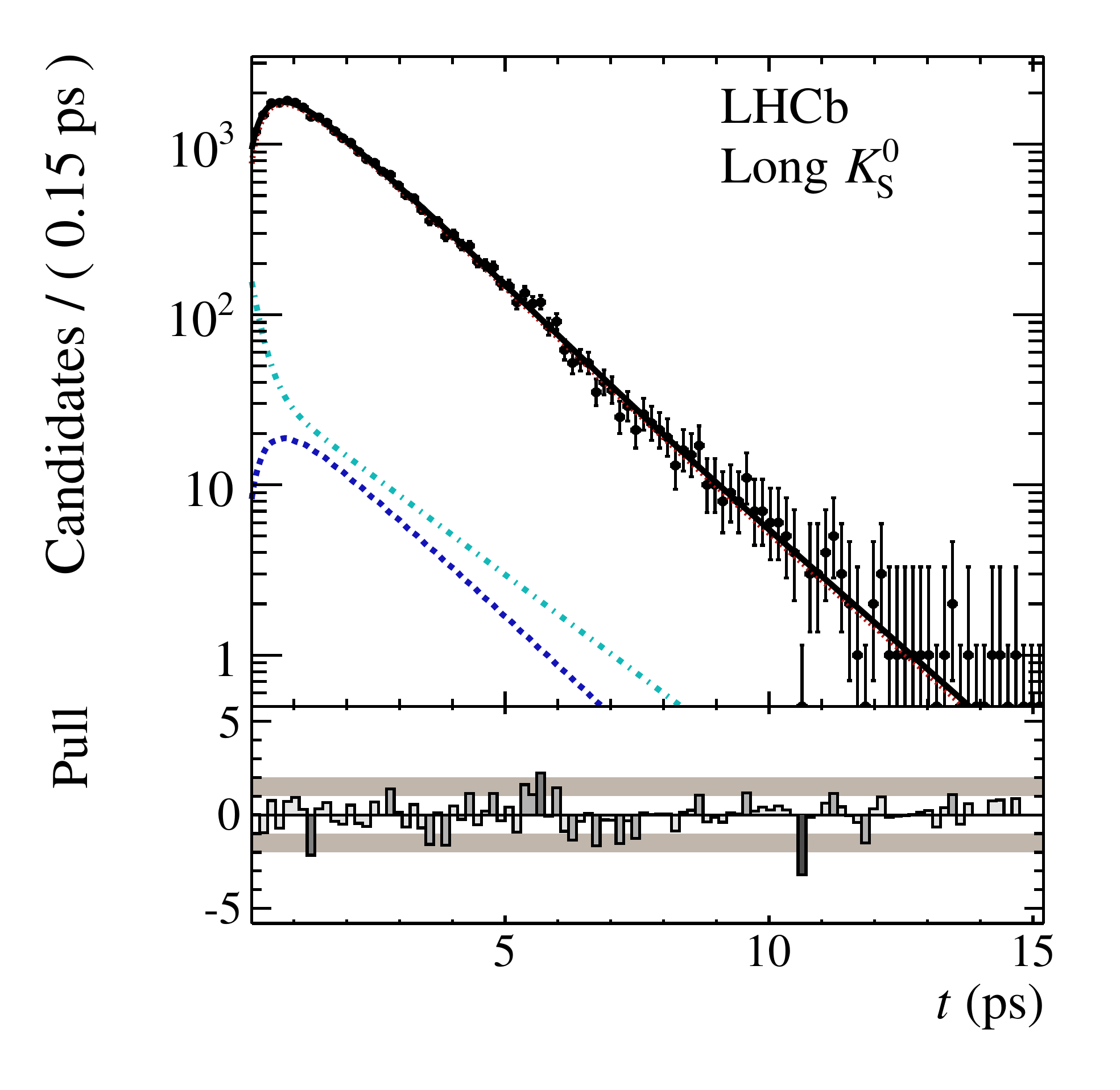}
\hfill
\includegraphics[width=0.49\textwidth]{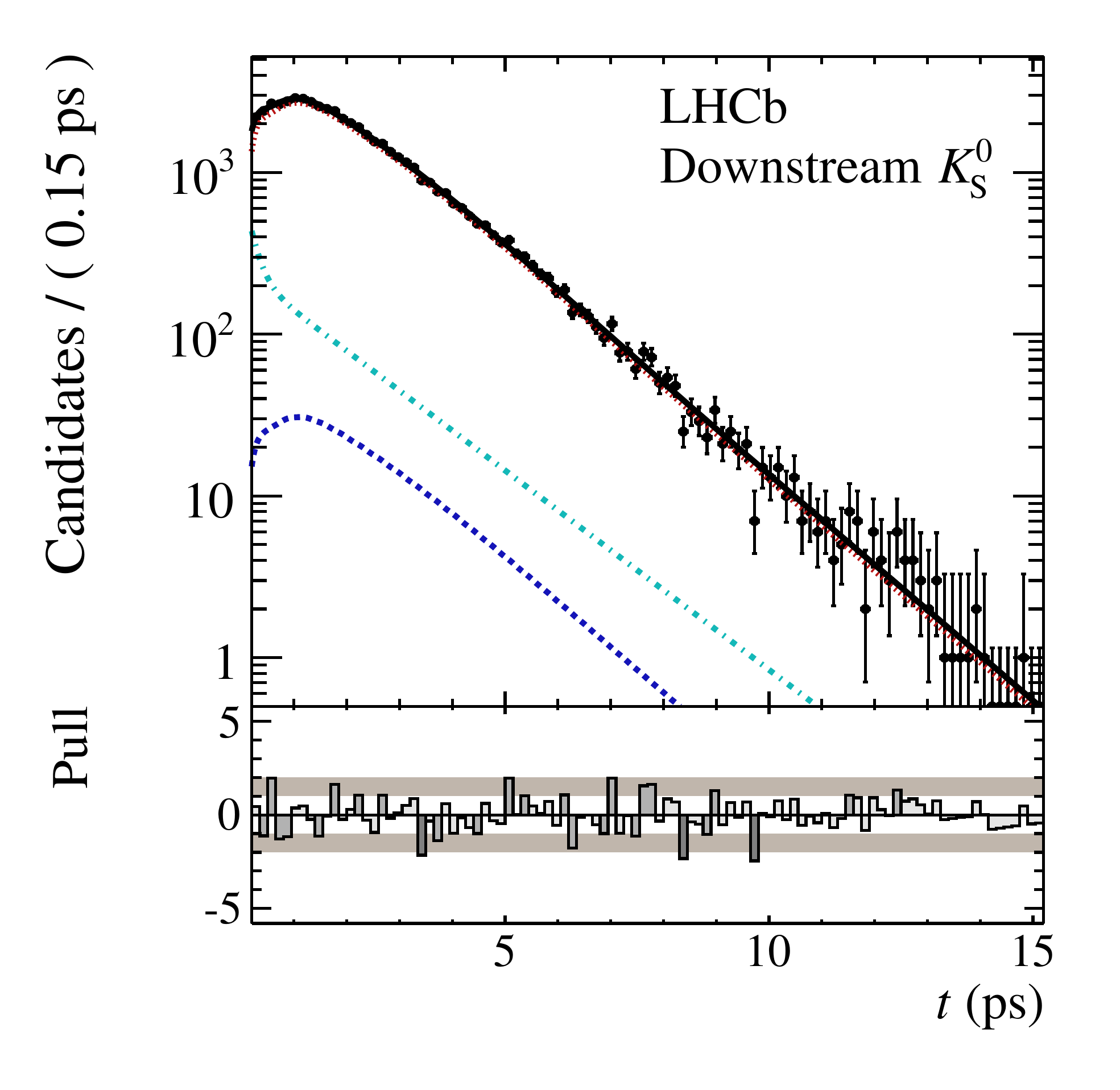}
\caption{Decay time distribution of $\B$ candidates in the (left) long \KS and (right) downstream \KS sample.
The fit projection is shown as solid black line.
Shown components are \BsToJPsiKS (dark blue, dashed), \BdToJPsiKS (red, dotted) and combinatorial background (turquoise, dash-dotted).}
\label{Fig:Time}
\end{figure}
%%%%%%%%%%%%%%%%%%%%%%%%%

%%%%%%%%%%%%%%%%%%%%%%%%%
\subsection{Likelihood fit}
The results are obtained from a simultaneous fit of the long and downstream \KS samples, using both the OS and SSK tagging information.
In addition to the five \CP{} observables, the nuisance parameters describing the mass (9 parameters), acceptance (12), background decay time (6) and event yields (18) are floated in the fit.
The observables $\Delta m_d$, $\tau_{\Bd}$, $\Delta m_s$, $\tau_{\Bs}$ and $\Delta\Gamma_s$, parametrising the \Bd and \Bs systems, and the effective \PB production asymmetries $A_{\text{prod}}(\Bd)$ and $A_{\text{prod}}(\Bs)$ of the long and downstream \KS samples are constrained using Gaussian functions.
The production asymmetries are defined in terms of the \B production cross-section $\sigma(\B)$ as $A_{\text{prod}}(\B) \equiv (\sigma(\Bbar) - \sigma(\B))/(\sigma(\Bbar) + \sigma(\B))$.
The statistical and systematic uncertainties on the constrained parameters are added in quadrature and treated together; the correlation $\rho(\Gamma_s, \Delta\Gamma_s) = -0.271$ \cite{HFAG} between the decay width and decay width difference of the \Bs meson is also included.
The effective \PB production asymmetries, specific to the data sample used in this analysis, are obtained by reweighting the results binned in \PB transverse momentum and pseudorapidity given in Ref.~\cite{LHCb-PAPER-2014-042}.
The obtained values are listed in Table \ref{Tab:ProdAsym}.

%%%%%%%%%%%%%%%%%%%%%%%%%
\begin{table}
\center
\caption{Effective \B production asymmetries specific to the data sample used in this analysis.}
\label{Tab:ProdAsym}
\begin{tabular}{lcc}
Sample & Mode & Value\\
\midrule
Long \KS & \Bd & $-0.0117 \pm 0.0057\:(\text{stat}) \pm 0.0013\:(\text{syst})$\\
Downstream \KS & \Bd & $-0.0095 \pm 0.0051\:(\text{stat}) \pm 0.0013\:(\text{syst})$\\
Long \KS & \Bs & $-0.041\phantom{0} \pm 0.032\phantom{0}\:(\text{stat}) \pm 0.003\phantom{0}\:(\text{syst})$\\
Downstream \KS & \Bs & $-0.022\phantom{0} \pm 0.024\phantom{0}\:(\text{stat}) \pm 0.003\phantom{0}\:(\text{syst})$
\end{tabular}
\end{table}
%%%%%%%%%%%%%%%%%%%%%%%%%

%%% Fit Validation
The likelihood fit is cross-checked using two independent implementations, and is validated with large sets of pseudoexperiments to thoroughly test several aspects of the analysis.
These also include the use of stand-alone event generators that produce samples independently of the fit implementations.
In addition, the fit model is tested on simulated data, with signal only and with both signal and background components present.
The results from the fit to the full data sample are compared to those from various subsamples, and to those obtained from a weighted fit to the $\BsToJPsiKS$ candidates only.
All tests agree with the expectations and no biases in the fit are found.

%%%%%%%%%%%%%%%%%%%%%%%%%
\subsection{Fit results}\label{Sec:FitResults}

The results of the $\BdToJPsiKS$ \CP asymmetries are
\begin{align*}
\cdir\left(\BdToJPsiKS\right)  & = \cdirBdVal \cdirBdStat\:,\\
\smix\left(\BdToJPsiKS\right)  & = \phantom{-}\smixBdVal \smixBdStat\:,
\end{align*}
where the uncertainties are statistical only.
They are compatible with the BaBar \cite{Babar:2009yr}, Belle \cite{Adachi:2012et} and latest LHCb \cite{sin2beta} results.
The results of the $\BsToJPsiKS$ \CP asymmetries are
\begin{align*}
\ADG\left(\BsToJPsiKS\right)  & = \ADGVal \ADGStat\:,\\
\cdir\left(\BsToJPsiKS\right)  & = \cdirVal \cdirStat\:,\\
\smix\left(\BsToJPsiKS\right)  & = \smixVal \smixStat\:,
\end{align*}
where the uncertainties are statistical only, and the observed event yields are summarised in Table \ref{Tab:Yields}.
The fit projections for the mass and decay time distributions are shown in Figs.~\ref{Fig:Mass} and \ref{Fig:Time}, respectively.
The statistical correlations between the $\BsToJPsiKS$ \CP observables are $\rho(\ADG, \cdir) = -0.07$, $\rho(\ADG, \smix) = -0.01$ and $\rho(\cdir, \smix) = -0.06$.
In addition, there is a $\mathcal{O}(10\%)$ correlation between $\ADG$ and the average decay width $\Gamma_s$ and decay width difference $\DGs$, and a $\mathcal{O}(10\%)$ correlation between $\smix$ and the $\Bs$ production asymmetries.
The confidence intervals for the three $\BsToJPsiKS$ \CP asymmetries are also calculated with the Feldman--Cousins method \cite{Feldman:1997qc,Karbach:2011uz}, which gives consistent results with the point estimates given above.

%%%%%%%%%%%%%%%%%%%%%%%%%
\begin{table}[ht]
\center
\caption{Fitted yields from the unbinned maximum likelihood fit.
The uncertainties are statistical only.}
\label{Tab:Yields}
\begin{tabular}{lrr}
\toprule
Yield & ~\hfill Long \KS\hfill~ & Downstream \KS\\
\midrule
\BdToJPsiKS & $27\:801 \pm 168$ & $51\:351 \pm 231$\\
\BsToJPsiKS & $307 \pm 20\phantom{0}$ & $601 \pm 30\phantom{0}$\\
Combinatorial background & $658 \pm 37\phantom{0}$ & $2\:852 \pm 74\phantom{0}$\\
\bottomrule
\end{tabular}
\end{table}
%%%%%%%%%%%%%%%%%%%%%%%%%

%!TEX root = main.tex

%%%%%%%%%%%%%%%%%%%%%%%%%%%%%%%%%%%%%%%%%%%%%%%%%%%%%%%%%%
\section{Systematic uncertainties}\label{Sec:Syst}
%%%%%%%%%%%%%%%%%%%%%%%%%%%%%%%%%%%%%%%%%%%%%%%%%%%%%%%%%%

%%% Modelling
A number of systematic uncertainties affecting the determination of the $\BsToJPsiKS$ \CP observables and the ratio of event yields $R$ are considered.
The main sources of systematic uncertainty are due to assumptions for modelling the different components of the multivariate PDF.
These uncertainties are estimated using large sets of simulated pseudoexperiments, in which the shapes and parameters of the individual PDF components are varied.
In the generation of the pseudoexperiments, the values of the parameters are fixed to the ones obtained in the fit to the data.
For each individual pseudoexperiment, the fitted values of the \CP observables and event yields are compared between the nominal fit and an alternative fit in which some of the shapes or nuisance parameters are varied.
The resulting differences between the fit values form a Gaussian-like distribution.
The mean and width of this distribution are added in quadrature and assigned as a systematic uncertainty.

Following this strategy, the systematic uncertainty due to the chosen mass model is evaluated by varying the Hypatia tail parameters within their uncertainties, replacing the signal model with a double Crystal Ball function \cite{Skwarnicki:1986xj}, and replacing the background model with a second-order Chebychev polynomial.
The latter variation has the largest impact on the \CP observables and yield ratio, and is used to assign a systematic uncertainty.

The systematic uncertainty associated with the decay time resolution is evaluated by varying the dilution of the resolution model, through changes of the resolution parameters, and by comparing the nominal model with one that includes a scale offset in the calibration functions for the per-candidate decay time uncertainty estimates.
The largest impact on the \CP observables and yield ratio originates from the limited knowledge on the decay time resolution of the long \KS sample.
This forms the dominant systematic uncertainty to the $\BsToJPsiKS$ \CP observables.

Systematic effects due to the modelling of the decay time acceptance mainly affect \ADG, and are evaluated by varying the empirical model for $\mathcal{E}(t)$.

The systematic uncertainty associated with the tagging calibration is obtained by comparing the nominal calibration with the largest and smallest effective tagging efficiency that can be obtained through changes of the calibration parameters within their respective uncertainties.

The mass resolution is assumed to be identical for the \Bd and \Bs signal modes, but it could depend on the mass of the reconstructed \B candidate.
This effect is studied by multiplying the width of the \Bs mass PDF by different scale factors, obtained by comparing \Bd and \Bs signal shapes in simulation.
These variations mainly affect the ratio of event yields.

Finally, a correlation between the reconstructed $B$ mass and decay time resolution is observed in simulated data.
The impact of neglecting this correlation in the fit to data is also evaluated with the simulated experiments.

The total systematic uncertainty and its sources are summarised in Table \ref{Tab:Syst_Summary}.

%%%%%%%%%%%%%%%%%%%%%%%%%%%%%%%%%%%%%%%%%%%%%%%%%%%%%%%%%%
\begin{table}[t]
\center
\caption{Summary of systematic uncertainties.}
\label{Tab:Syst_Summary}
\begin{tabular}{lrrrcc}
\toprule
& & & & \multicolumn{1}{c}{Long} & \multicolumn{1}{c}{Downstream} \\
Source
  & \multicolumn{1}{c}{$\ADG$}
  & \multicolumn{1}{c}{$\cdir$}
  & \multicolumn{1}{c}{$\smix$}
  & \multicolumn{1}{c}{$R\times 10^{5}$}
  & \multicolumn{1}{c}{$R\times 10^{5}$} \\
\midrule
Mass modelling         & 0.045 & 0.009 & 0.009 & 15.5 & 17.2 \\
Decay-time resolution  & 0.038 & 0.066 & 0.070 & \phantom{1}0.6 & \phantom{1}0.3 \\
Decay-time acceptance  & 0.022 & 0.004 & 0.004 & \phantom{1}0.6 & \phantom{1}0.5 \\
Tagging calibration    & 0.002 & 0.021 & 0.023 & \phantom{1}0.1 & \phantom{1}0.2 \\
Mass resolution        & 0.010 & 0.005 & 0.006 & 12.6 & \phantom{1}8.0 \\
Mass--time correlation  & 0.003 & 0.037 & 0.036 & \phantom{1}0.2 & \phantom{1}0.1 \\
\midrule
Total                  & 0.064 & 0.079 & 0.083 & 20.0 & 19.0 \\
                                                  
\bottomrule
\end{tabular}
\end{table}
%%%%%%%%%%%%%%%%%%%%%%%%%%%%%%%%%%%%%%%%%%%%%%%%%%%%%%%%%%

%!TEX root = main.tex

%%%%%%%%%%%%%%%%%%%%%%%%%%%%%%%%%%%%%%%%%%%%%%%%%%%%%%%%%%
\section{Branching ratio measurement}\label{Sec:BR}
%%%%%%%%%%%%%%%%%%%%%%%%%%%%%%%%%%%%%%%%%%%%%%%%%%%%%%%%%%

%%% Main Formula
The measured ratio of branching fractions is calculated from the event yields using Eq.~\eqref{Eq:BR_Def}.
%%% Selection Efficiency
The selection efficiencies and their ratio $f_{\text{sel}}$ are evaluated using simulated data.
As the simulated data are generated with different values for the lifetime $\tau_{\Bs}$, decay width difference $\DGs$ and acceptance parameters compared to those measured in the collision data, correction factors are applied.
This leads to a ratio of total selection efficiencies of $f_{\text{sel}} = \SelLL$ for the long \KS sample and $f_{\text{sel}} = \SelDD$ for the downstream \KS samples.

%%% Main Results
Combining the results in Table \ref{Tab:Yields} with the systematic uncertainties in Table \ref{Tab:Syst_Summary} yields
\begin{align*}
R\:\text{(long)} & = \rawLL\:,\\
R\:\text{(downstream)} & = \rawDD
\end{align*}
for the long and downstream \KS samples, respectively.
A weighted average of the combinations $R\times f_{\text{sel}}$ for the long and downstream \KS samples is performed, assuming that they are uncorrelated measurements.
The measured ratio of branching fractions is then given by
\begin{equation*}
\frac{\BR(\BsToJPsiKS)}{\BR(\BdToJPsiKS)} = \BRBsBdVal \BRBsBdStat \BRBsBdSyst \BRBsBdfds\:.
\end{equation*}
where the third uncertainty is due to the uncertainty in $f_s/f_d$.

Combining the ratio of branching fractions with the known $\decay{\Bd}{\jpsi{}\Kz}$ branching fraction \mbox{$\BR(\decay{\Bd}{\jpsi{}\Kz}) = (8.97 \pm 0.35)\times 10^{-4}$} \cite{PDG2014}, which accounts for the difference in production rates for the \Bu{}\Bub and \Bd{}\Bdb pairs at the \FourS{} resonance, i.e.\ $\Gamma(\Bu{}\Bub)/\Gamma(\Bd{}\Bdb) = 1.058 \pm 0.024$ \cite{HFAG}, the $\BsToJPsiKS$ branching fraction is
\begin{align*}
& \BR(\BsToJPsiKS) = \\
& \left[\BRBsVal \BRBsStat \BRBsSyst \BRBsfds \BRBsPDG\right]\times10^{-5}\:,
\end{align*}
where the last uncertainty comes from the $\decay{\Bd}{\jpsi{}\Kz}$ branching fraction.

%%%%%%%%%%%%%%%%%%%%%%%%%%%%%%%%%%%%%%%%%%%%%%%%%%%%%%%%%%
\section{Conclusion}\label{Sec:Conclusion}
%%%%%%%%%%%%%%%%%%%%%%%%%%%%%%%%%%%%%%%%%%%%%%%%%%%%%%%%%%

This paper presents the first measurement of the time-dependent \CP{} violation observables in the decay $\BsToJPsiKS$ and an updated measurement of its time-integrated branching fraction.
Both measurements are performed using a data set corresponding to an integrated luminosity of 3.0\:\invfb of $pp$ collisions recorded by the LHCb detector at centre-of-mass energies of 7 and 8 \tev.

The results on the \CP observables are
\begin{align*}
\ADG\left(\BsToJPsiKS\right)  & = \ADGVal \ADGStat \ADGSyst\:,\\
\cdir\left(\BsToJPsiKS\right)  & = \cdirVal \cdirStat \cdirSyst\:,\\
\smix\left(\BsToJPsiKS\right)  & = \smixVal \smixStat \smixSyst\:.
\end{align*}
The large statistical uncertainties on these results do not allow for a conclusive comparison with the predictions in Eq.~\eqref{Eq:CPprediction} nor do they provide constraints on the shift parameter $\Delta\phid$ affecting \CP measurements in $\BdToJPsiKS$.

The ratio of time-integrated branching fractions is measured to be
\begin{equation*}
\frac{\BR(\BsToJPsiKS)}{\BR(\BdToJPsiKS)} = \BRBsBdVal \BRBsBdStat \BRBsBdSyst \BRBsBdfds\:.
\end{equation*}
This result is the single most precise measurement of this quantity, and supersedes the previous LHCb measurement \cite{LHCb-PAPER-2013-015}.

% Do not include this in analysis note and conference reports
\section*{Acknowledgements}
 
\noindent We express our gratitude to our colleagues in the CERN
accelerator departments for the excellent performance of the LHC. We
thank the technical and administrative staff at the LHCb
institutes. We acknowledge support from CERN and from the national
agencies: CAPES, CNPq, FAPERJ and FINEP (Brazil); NSFC (China);
CNRS/IN2P3 (France); BMBF, DFG, HGF and MPG (Germany); INFN (Italy); 
FOM and NWO (The Netherlands); MNiSW and NCN (Poland); MEN/IFA (Romania); 
MinES and FANO (Russia); MinECo (Spain); SNSF and SER (Switzerland); 
NASU (Ukraine); STFC (United Kingdom); NSF (USA).
The Tier1 computing centres are supported by IN2P3 (France), KIT and BMBF 
(Germany), INFN (Italy), NWO and SURF (The Netherlands), PIC (Spain), GridPP 
(United Kingdom).
We are indebted to the communities behind the multiple open 
source software packages on which we depend. We are also thankful for the 
computing resources and the access to software R\&D tools provided by Yandex LLC (Russia).
Individual groups or members have received support from 
EPLANET, Marie Sk\l{}odowska-Curie Actions and ERC (European Union), 
Conseil g\'{e}n\'{e}ral de Haute-Savoie, Labex ENIGMASS and OCEVU, 
R\'{e}gion Auvergne (France), RFBR (Russia), XuntaGal and GENCAT (Spain), Royal Society and Royal
Commission for the Exhibition of 1851 (United Kingdom).

\addcontentsline{toc}{section}{References}
\setboolean{inbibliography}{true}
\bibliographystyle{LHCb}
\bibliography{Bs2JpsiKs}

% Author List ----------------------------                                                                                                                                                                                                                                                                                                
\newpage
%%%%%%%%%%%%%%%%%%%%%%%%%%%%%%%%%%%%%%%%%%
\centerline{\large\bf LHCb collaboration}
\begin{flushleft}
\small
R.~Aaij$^{41}$, 
B.~Adeva$^{37}$, 
M.~Adinolfi$^{46}$, 
A.~Affolder$^{52}$, 
Z.~Ajaltouni$^{5}$, 
S.~Akar$^{6}$, 
J.~Albrecht$^{9}$, 
F.~Alessio$^{38}$, 
M.~Alexander$^{51}$, 
S.~Ali$^{41}$
% ...

{\footnotesize \it
% ...
$ ^{5}$Clermont Universit\'{e}, Universit\'{e} Blaise Pascal, CNRS/IN2P3, LPC, Clermont-Ferrand, France\\
$ ^{6}$CPPM, Aix-Marseille Universit\'{e}, CNRS/IN2P3, Marseille, France\\
% ...
$ ^{9}$Fakult\"{a}t Physik, Technische Universit\"{a}t Dortmund, Dortmund, Germany\\
% ...
$ ^{37}$Universidad de Santiago de Compostela, Santiago de Compostela, Spain\\
$ ^{38}$European Organization for Nuclear Research (CERN), Geneva, Switzerland\\
% ...
$ ^{41}$Nikhef National Institute for Subatomic Physics, Amsterdam, The Netherlands\\
% ...
$ ^{46}$H.H. Wills Physics Laboratory, University of Bristol, Bristol, United Kingdom\\
% ...
$ ^{51}$School of Physics and Astronomy, University of Glasgow, Glasgow, United Kingdom\\
$ ^{52}$Oliver Lodge Laboratory, University of Liverpool, Liverpool, United Kingdom\\
% ...
}
\end{flushleft}
%%%%%%%%%%%%%%%%%%%%%%%%%%%%%%%%%%%%%%%%%%

% This should be taken out in the final paper
%\clearpage
%\input{supplementary}

\end{document}